\documentclass[a4paper,fleqn,usenatbib]{mnras}

\usepackage[T1]{fontenc}
\usepackage{ae,aecompl}

\usepackage{graphicx}	
\usepackage{amsmath}	
\usepackage{amssymb}	
\usepackage{multirow}
\usepackage{tabularx}

\title[Maximum CR energy in Cygnus A primary hotspot]{On the maximum energy of
  non-thermal particles in the primary hotspot of Cygnus A}

\author[A. T. Araudo et al.]{
Anabella T. Araudo$^{1,2}$\thanks{E-mail: anabella.araudo@asu.cas.cz},
Anthony R. Bell$^{4}$,
Katherine M. Blundell$^{3}$, and 
\newauthor{James H. Matthews$^{3}$}
\\
$^{1}$Astronomical Institute of the Czech Academy of Sciences, Bocni II 1401, Prague, CZ-14100 Czech Republic  \\
$^{2}$Laboratoire Univers et Particules de Montpellier CNRS/Universite de Montpellier, Place E. Bataillon, 34095 Montpellier, France\\
$^{3}$University of Oxford, Astrophysics, Keble Road, 
Oxford OX1 3RH, UK\\
$^{4}$University of Oxford, Clarendon Laboratory, Parks Road, 
Oxford OX1 3PU, UK
}

\date{Accepted XXX. Received YYY; in original form ZZZ}

\pubyear{2017}

\begin{document}
\label{firstpage}
\pagerange{\pageref{firstpage}--\pageref{lastpage}}
\maketitle

\begin{abstract}
We study particle acceleration and magnetic field amplification in the 
primary hotspot in the northwest jet of radiogalaxy Cygnus~A. By using the 
observed flux density at 43~GHz
in a well resolved region of this hotspot, we determine the minimum value
of the jet density and constrain the magnitude of the magnetic field. 
We find that a jet with density greater than $5\times10^{-5}$~cm$^{-3}$ and 
hotspot magnetic field in the range $50-400$~$\mu$G are 
required to explain the synchrotron emission at 43~GHz.
The upper-energy cut-off in the hotspot synchrotron spectrum is at a frequency 
$\lesssim 5\times10^{14}$~Hz,
indicating that the maximum energy of non-thermal electrons accelerated
at the jet reverse shock is $E_{e,\rm max}\sim 0.8$~TeV in a magnetic field of 
100~$\mu$G. 
Based on the condition that the magnetic-turbulence scale length has to be 
larger than the plasma
skin depth, and that the energy density in non-thermal particles cannot 
violate the limit imposed by the 
jet kinetic luminosity, we show that $E_{e,\rm max}$ cannot be 
constrained by synchrotron losses as traditionally assumed. 
In addition to that, and assuming that the shock is quasi-perpendicular,
we show that non-resonant hybrid instabilities 
generated by the streaming of cosmic rays with energy  $E_{e,\rm max}$ can
grow fast enough 
to amplify the jet magnetic field up to $50-400$~$\mu$G and accelerate 
particles up to the maximum energy $E_{e,\rm max}$ observed in the Cygnus~A 
primary hotspot.
\end{abstract}


\section{Introduction}

Type II Fanaroff-Riley (FR) radiogalaxies exhibit well collimated jets with 
bright radio synchrotron knots (hotspots) at the termination region. 
Electrons radiating in the hotspot  are locally accelerated in the jet reverse 
shock, and they reach a maximum energy $E_{e,\rm max}$  inferred from the 
Infrared (IR)/optical cut-off frequency ($\nu_{\rm c}$) of the synchrotron  
spectrum:  
\begin{equation}
\frac{E_{e,\rm max}}{\rm TeV} \sim 0.8
\left(\frac{\nu_{\rm c}}{5\times10^{14}\,{\rm Hz}}\right)^{\frac{1}{2}}
\left(\frac{B}{100\,\mu{\rm G}}\right)^{-\frac{1}{2}},
\label{Ec}
\end{equation}
where $B$ is the magnetic field \cite[e.g.][]{3c273-Natur,Brunetti_03}. 
In some cases, X-rays are also detected and modeled as 
synchrotron self Compton emission
and Compton up-scattering of Cosmic Microwave Background photons
\cite[e.g.][]{Perlman_10,Wilson_00}. We note however that in very few cases
X-ray synchrotron emission is proposed \citep{Tingay_08,Orienti_17}.

Ions can also be accelerated in the jet reverse shock. Given that hadronic 
losses are very slow in
low density plasmas such as  the termination region of FR~II radiogalaxy jets, 
protons might achieve 
energies as large as the  limit imposed by the size of the system, usually 
called 
"Hillas limit" \citep{Hillas,Lagage-Cesarsky}. In particular,  
mildly relativistic  shocks with velocity  $v_{\rm sh}=c/3$ 
might accelerate particles with  Larmor radius
$r_{\rm g} \sim R_{\rm j}$, where $R_{\rm j}$ is the jet width at the
termination region. Particles with such a large $r_{\rm g}$  have energy 
\begin{equation}
\frac{E_{\rm Hillas}}{\rm EeV} \sim 100
\left(\frac{v_{\rm sh}}{c/3}\right)
\left(\frac{B}{100 \,\rm \mu G}\right)
\left(\frac{R_{\rm j}}{\rm kpc}\right),
\label{hillas}
\end{equation}
as expected for Ultra High Energy Cosmic Rays (UHECRs) 
\cite[e.g.][]{Rachen_93,Norman_95}. \cite{Tony_17} examine 
the maximum energy to which Cosmic Rays (CR) can be accelerated by
relativistic shocks, showing that acceleration of protons to 100~EeV is 
unlikely.
\cite[See also][]{Kirk_Brian_10,lemoine-pelletier-10,Sironi_13,Brian_14}.

In \cite{cutoff_nrh} we have shown that hotspots of FR~II radiogalaxies are 
very poor accelerators. We have shown that the maximum energy of
non-thermal electrons accelerated at the reverse shocks is not determined 
by synchrotron losses, unless very extreme conditions in
the plasma are assumed\footnote{In our previous papers we called 
$E_{\rm c} = E_{e,\rm max}$ and $E_{\rm uhecr}=E_{\rm Hillas}$.}. 
By equating the acceleration and synchrotron cooling timescales,  we show that
the mean free path of the most energetic electrons accelerated at the 
jet termination shocks is greater than the maximum value 
imposed by plasma physics for canonical values of the magnetic field and
jet density. We demonstrated this by 
considering the sample of 8 hotspots observed with high spatial
resolution at optical, IR and radio wavelengths by \cite{Mack_09}. 

If synchrotron losses do not balance energy gain, the electrons' maximum 
energy $E_{e,\rm max}$ is ultimately
determined by the ability to scatter particles in the shock environment, and 
this limit applies to both electrons and protons. Assuming that the jet
magnetic field downstream of the shock  is quasi-perpendicular, 
we found  that non-resonant (Bell) turbulence generated by the streaming of 
CRs  can grow fast enough 
to amplify the jet magnetic field by about two orders of magnitude
and accelerate particles up to $E_{e,\rm max}\sim 0.1-1$~TeV.

In the present paper we study the FR~II radiogalaxy Cygnus A, having
a redshift $z\sim 0.05607$ ($d \sim$227.3~Mpc, where $d$ is the distance 
from Earth) in the Cygnus galaxy cluster \citep{Owen_97}.  
The northwest jet terminates at $\sim 60$~kpc
from the central source where the primary (B) and secondary (A) hotspots 
are detected
\footnote{The northwest primary and secondary hotspots are sometimes called B and A,
respectively \cite[e.g.][]{Stawarz_07}.}. 
\cite{Stawarz_07} modelled the radio-to-X-rays non-thermal 
emission from  the secondary
hotspots in the one-zone approximation and assumed that $E_{e,\rm max}$ is 
determined by synchrotron cooling. In this work we  apply the same 
methodology presented in 
\cite{cutoff_nrh} to the northwest primary hotspot. We improve our previous 
model by removing the assumption that the jet density (at the termination 
region) is $n_{\rm jet}=10^{-4}$~cm$^{-3}$.
Using the 43~GHz high spatial resolution data we constrain the magnetic field
and the jet density (Section~\ref{B_43}). On the other hand, using the 
cut-off of the synchrotron spectrum  determined from IR and optical
emission we show that $E_{e,\rm max}$ cannot be determined by synchrotron 
cooling, unless the jet density is of the order of the 
density in the external medium (Section~\ref{cutoff}). 
Finally, assuming that the magnetic field downstream of the shock is 
quasi-perpendicular, we constrain the scale size of magnetic turbulence
(Section~\ref{s_optical}) and show that it can be excited through 
the non resonant hybrid (NRH) instability (Section~\ref{nrh}). 
We conclude that the primary hotspot in Cygnus~A is a clear example where 
particle acceleration
is not constrained by synchrotron losses. 
Throughout this paper we use cgs units and the 
cosmology $H_0 = 71$~km~s$^{-1}$~Mpc$^{-1}$, $\Omega_0 = 1$ and 
$\Lambda_0 =  0.73$. One arcsecond represents $1.044$~kpc on the plane of 
the sky at $z = 0.05607$.

\begin{figure}
\includegraphics[angle=270,width=1.3\columnwidth]{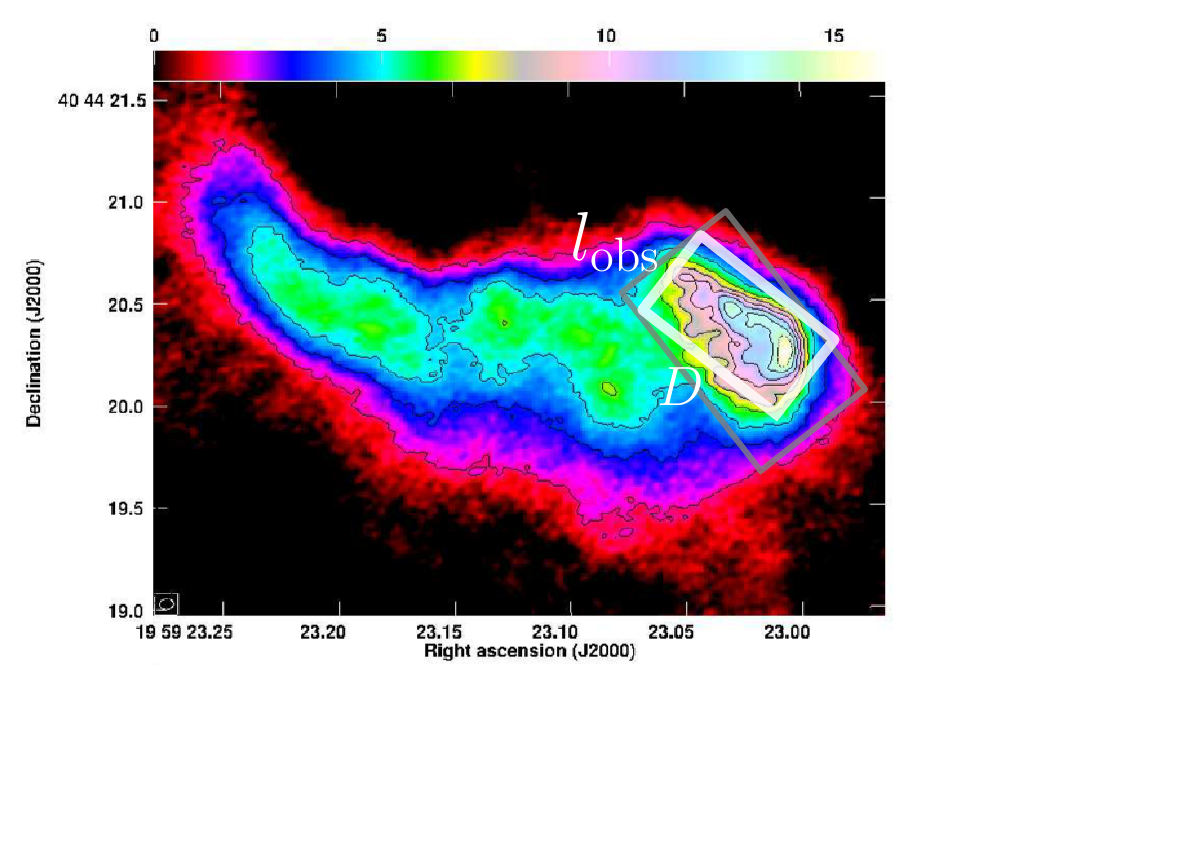}
\vspace{-0.8cm}
\caption{Cygnus~A primary hotspot at 43~GHz. The grey rectangle indicates
region considered in Pyrzas et al. (2015) to compute the spectral index,
whereas the white rectangle of size $l_{\rm obs} = 0.5^{\prime\prime}$ and 
$D = 0.9^{\prime\prime}$ is the region considered in our study to compute the
magnetic field in the hotspot. It is approximately drawn to match the 
half height points of the emission profile.
Adapted from Pyrzas et al. (2015).}
\label{primary_43}
\end{figure}

\section{Synchrotron radio emission from the northwest primary hotspot}
\label{primary_HS}

The northwest primary hotspot has been detected with the MERLIN interferometer
at 151~MHz \citep{Leahy_89} and with the
Very Large Array (VLA) at frequencies from 327~MHz to
87~GHz \cite[e.g.][]{Carilli_91}. 
In addition to that, 230~GHz emission was detected with the
BIMA array with $1^{\prime\prime}$ angular resolution.
\cite{Wright_04} made spectral index maps and found that the $5-230$,
$5-15$, and $15-230$~GHz spectral indices are 
$\alpha_5^{230} =\alpha_5^{15} = \alpha_{15}^{230} =1.13$, whereas
$\alpha_{87}^{230} =1.23$.
Therefore no spectral break is observed between 5 and 230~GHz. 
These steep radio spectral indices would indicate
that electrons emitting synchrotron radiation at these frequencies 
radiate most of their energy in the hotspot. 
However, recent analysis from the same set of VLA data  shows that the 
spectral index from 5 to 43~GHz is $\alpha_5^{43} \sim 0.72$ \citep{Pyrzas_15},
consistent with standard diffusive shock acceleration in the slow cooling 
regime. In the following section we will consider the  well resolved 
emission at 43~GHz to constrain the value of the magnetic 
field\footnote{The VLA beam-size at 43~GHz is 0.07$\times$0.06~arcsec$^2$.}.

\subsection{Constraining the magnetic field with the synchrotron emission 
at 43~GHz}
\label{B_43}

Figure~\ref{primary_43} shows the hotspot at 43~GHz, 
where the region considered by \cite{Pyrzas_15} to calculate the spectral
index ($\alpha_5^{43}$) is indicated by the grey rectangle of 
0.7$\times$1.2~arcsec$^2$. For our study, we select a 
region of $0.5\times 0.9$~arcsec$^2$ (indicated by the 
white rectangle in Figure~\ref{primary_43}) defined by the
half-height points of the emission peak.
Considering that the radio emitter is a cylinder
of diameter $D=0.9^{\prime\prime}$ and width (projected in the plane of the sky) 
$l_{\rm obs}=0.5^{\prime\prime}$,  the emitter volume 
at 43~GHz is $V = \pi D^2l_{\rm obs}/4 \sim  0.32$~arcsec$^3$ 
(i.e. $V \sim 0.36$~kpc$^3$).
The background emission corrected flux at 43~GHz is $f_{43}=0.36$~Jy
\citep{Pyrzas_15}, and the specific luminosity is
$L_{43} = 43\times10^9 f_{43} 4 \pi d^2\sim 9\times10^{41}$~erg~s$^{-1}$.
We model the synchrotron radio emission  in $V$ as produced 
by non-thermal electrons following  a power-law 
energy distribution $N_e \propto E_e^{-p}$, with  $p = 2\alpha_5^{43} +1 = 2.44$
and $E_e \ge E_{e,\rm min} = m_ec^2 \gamma_{e,\rm min}$, where
\begin{equation}
\gamma_{e,\rm min} \sim 450
\left(\frac{\nu_{\rm min}}{0.1\,\rm GHz}\right)^{\frac{1}{2}}
\left(\frac{B}{100\,\mu{\rm G}}\right)^{-\frac{1}{2}}
\label{gamma_c}
\end{equation}
and $\nu_{\rm min}$ is the frequency of the low-energy turnover\footnote{
Using the Low Frequency Array (LOFAR) between 109 and 183 MHz, at an angular
resolution of $\sim 3.5^{\prime\prime}$, \cite{CygA_gammamin} found that
the low energy turnover of the secondary hotspots synchrotron spectra
in Cygnus~A is at $\sim 150$~MHz.} \cite[e.g.][]{CygA_gammamin}. 
We insert  $\gamma_{e,\rm min}$ and the numerical values of $V$, $p$, $\nu$, and
$L_{43}$ (see Table~\ref{tab}) in equations~20 and 21 in \cite{cutoff_nrh}. 
We find that  the energy density 
in non-thermal electrons determined from the synchrotron emission at 43~GHz is
\begin{equation}\label{Ue}
\frac{U_{e}}{\rm erg\,cm^{-3}}   \sim  2\times10^{-8}
\left(\frac{\nu_{\rm min}}{{0.1\,\rm GHz}}\right)^{-0.22}
\left(\frac{B}{100\,\rm \mu G}\right)^{-\frac{3}{2}}
\left(\frac{V}{0.36\,\rm kpc^3}\right)^{-1},
\end{equation}
and the magnetic field in equipartition with non-thermal electrons and 
protons would be
\begin{equation}\label{Beq}
\frac{B_{\rm eq}}{\rm \mu G}  \sim 390
\left(\frac{1+a}{2}\right)^{\frac{2}{7}}
\left(\frac{\nu_{\rm min}}{0.1\rm GHz}\right)^{-0.06}
\left(\frac{V}{0.36\,\rm kpc^3}\right)^{-\frac{2}{7}},
\end{equation}
where the energy density in non-thermal protons is $U_p = a U_e$ and
therefore the non-thermal energy density is $U_{\rm nt} = (1+a)U_e$. 
Note the weak dependence of $U_e$ and $B_{\rm eq}$ on $\nu_{\rm min}$.
We keep $V$ fixed in Eqs.~(\ref{Ue}) and (\ref{Beq}).

\begin{figure}
\includegraphics[width=0.49\textwidth]{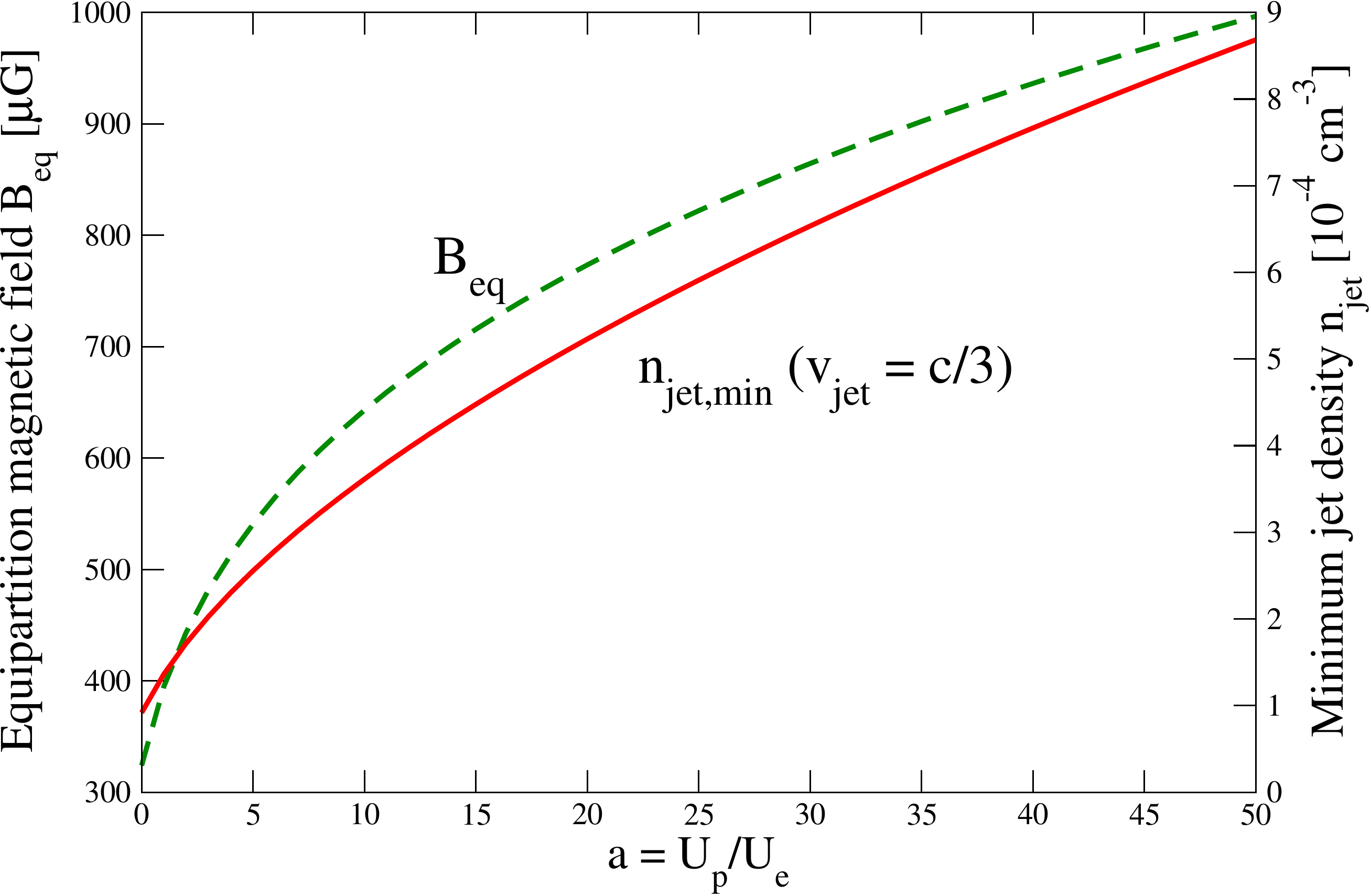}
\caption{\emph{Left axis:} Magnetic field in equipartition with non-thermal 
particles: $B_{\rm eq}^2/(8\pi) = (1+a)U_e$ (green-dashed line). 
\emph{Right axis:} Minimum jet matter density ($n_{\rm jet,min}$) considering  
$v_{\rm jet} = c/3$ (red-solid line).}
\label{B_a}
\end{figure}

The jets of Cygnus A suggest a precession pattern from which the jet velocity 
was estimated as $0.2c < v_{\rm jet}< 0.5c$ in the termination region
\cite[see][and references therein]{Steenbrugge_08}. 
The jet kinetic energy density is
\begin{equation}\label{Ukin}
\frac{U_{\rm kin}}{\rm erg\,cm^{-3}}  \sim 9\times 10^{-9}
\left(\frac{n_{\rm jet}}{10^{-4}\,\rm cm^{-3}}\right)
\left(\frac{\Gamma_{\rm jet} -1}{0.06}\right),
\end{equation}
where $\Gamma_{\rm jet} = 1.06$ is the jet bulk Lorentz factor when 
$v_{\rm jet} = c/3$ (see Table~\ref{tab}).
By setting the extreme condition $U_{\rm kin}= 2 B_{\rm eq}^2/(8\pi)$, 
i.e. all the jet kinetic energy density in the shock upstream region is
converted into magnetic ($U_{\rm mag}$) and non-thermal ($U_{\rm nt}$) energy 
densities in the downstream region (the hotspot) and that 
$U_{\rm mag}=U_{\rm nt}=(1+a)U_e$ (the equipartition condition),
the minimum jet matter density (at the termination region) is
\begin{eqnarray}\label{n_jet_min}
\begin{aligned}
n_{\rm jet,min}& = 2\left(\frac{B_{\rm eq}^2}{8\pi}\right)
\left(\frac{1}{m_pc^2(\Gamma_{\rm jet} -1)}\right)\\
&\sim 1.36\times10^{-4}
\left(\frac{1+a}{2}\right)^{\frac{4}{7}}
\left(\frac{\nu_{\rm min}}{0.1\rm GHz}\right)^{-0.12}\\
&\times \left(\frac{\Gamma_{\rm jet} -1}{0.06}\right)^{-1}
\left(\frac{V}{0.36\,\rm kpc^3}\right)^{-\frac{4}{7}}\,{\rm cm^{-3}}.
\end{aligned}
\end{eqnarray}
In Figure~\ref{B_a} we  plot 
$n_{\rm jet,min}$ for $0\le a \le 50$ and considering $v_{\rm jet} = c/3$. 
Note that  $n_{\rm jet,min} \propto 1/(\Gamma_{\rm jet}-1)$ and therefore
it is $\sim$3.7 times larger and 0.41 times smaller than the values plotted in 
Figure~\ref{B_a} when $v_{\rm jet} = c/5$ and $c/2$, respectively. In
Table~\ref{tab} we list $n_{\rm jet,min}$ for  $v_{\rm jet} = c/5$, $c/3$, and
$c/2$, and $a=1$.
In relativistic shocks we do not expect $a$ much larger than 1, 
and  hereafter we  consider $a=1$. Therefore, the energy density in 
non-thermal particles is $U_{\rm nt} = 2 U_e$. 

\begin{table}
\caption{Physical parameters of the northwest primary hotspot.}
\centering
\label{tab}
\begin{tabular}{l|l|l}
\hline
$z=0.05607$ & $d=227.3$~Mpc & $1^{"}=1.044$~kpc\\
$\alpha = 0.72$ & $p=2.44$ &\\
$\nu_{\rm min}= 0.1$~GHz & $\nu=43$~GHz & $\nu_{\rm c}=5\times10^{14}$~Hz \\
\hline
\multicolumn{3}{c}{43 GHz}\\
\hline
$l_{\rm obs}=0.5^{"}$ & $D=0.9^{"}$ & $V= 0.36$~kpc$^3$\\
$f_{43} = 0.36$~Jy & $L_{43}=9\times 10^{41}$~erg~s$^{-1}$ & $B_{\rm eq} =
390$~$\mu$G ($a=1$)\\
\hline\hline
$v_{\rm jet}$ & $\Gamma_{\rm jet}$& $n_{\rm jet,min}$ [cm$^{-3}$]\\
\hline
$c/2$ & $1.15$& $5.44\times10^{-5}$ ($a=1$)\\
$c/3$ & $1.06$& $1.36\times10^{-4}$ ($a=1$)\\
$c/5$ & $1.02$& $4.08\times10^{-4}$ ($a=1$)\\
\hline\hline
\end{tabular}
\end{table}

\begin{figure}
\includegraphics[width=0.49\textwidth]{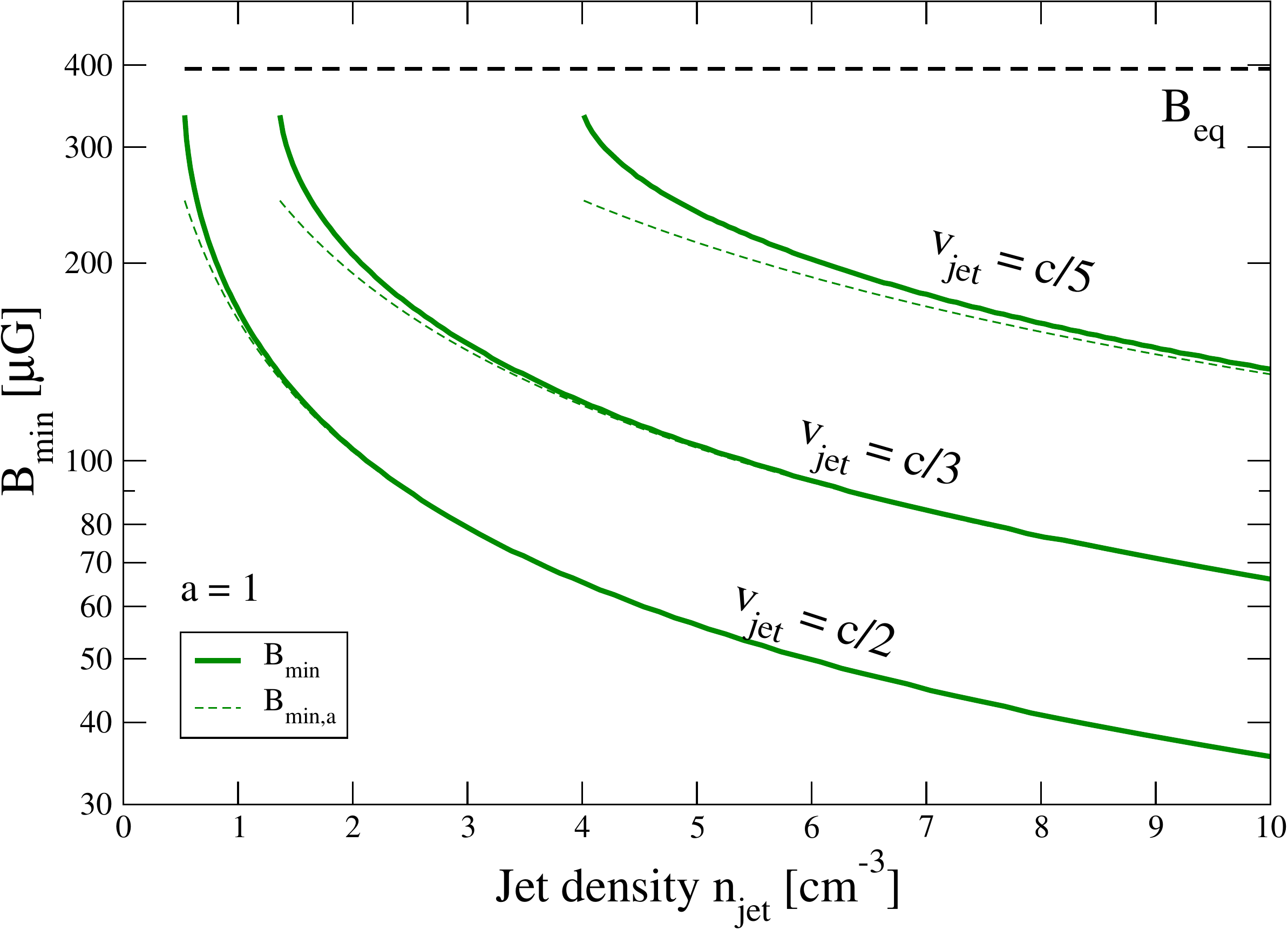}
\caption{Minimum magnetic field required to explain the flux at 43~GHz. 
$B_{\rm min}$ (green-solid lines) and $B_{\rm min,a}$  (green-dashed lines)
are plotted for the case of $a=1$ and $v_{\rm jet} = c/2$, $c/3$ and $c/5$. 
The equipartition magnetic field is also indicated (black-dashed line).
\label{B}}
\end{figure}

Given that $2 U_e > B^2/(8\pi)$ when $n_{\rm jet} > n_{\rm jet,min}$,  most of the 
jet kinetic energy goes to non-thermal  particles when we consider that there 
is only magnetic and non-thermal energy in the hotspot. Therefore, we find 
the magnetic field minimum value ($B_{\rm min}$) required to explain the 
emission at 43~GHz by setting the condition 
$U_{\rm kin} = 2 U_e + B_{\rm min}^2/(8\pi)$.  In Figure~\ref{B} we plot 
$B_{\rm min}$ (green-solid line) for $v_{\rm jet}=c/2$, $c/3$, and $c/5$, and 
from $n_{\rm jet,min}$ to $10^{-3}$~cm$^{-3}\sim 0.1\,n_{\rm ext}$, where 
$n_{\rm ext} \sim 10^{-2}$~cm$^{-3}$ is the density in the external medium 
\cite[e.g.][]{Wilson_06}. Values 
of $n_{\rm jet}$ larger than $10^{-3}$~cm$^{-3}$
would be very unrealistic given that the jet to external medium density ratio 
in adiabatic flows
is expected to be $\sim 10^{-2}$. In fact, \cite{polarization} found that the
plasma density in the jet of Cygnus~A is smaller than $4\times10^{-4}$~cm$^{-3}$.
In order to provide
an analytical expression ($B_{\rm min,a}$) of $B_{\rm min}$ we set the unrealistic 
condition $2U_e = U_{\rm kin}$ and therefore 
\begin{eqnarray}\label{Bmin}
\begin{aligned}
\frac{B_{\rm min,a}}{\rm \mu G}   & \sim  305
\left(\frac{\nu_{\rm min}}{0.1\,{\rm GHz}}\right)^{-0.15}\\
&\times\left[ \left(\frac{n_{\rm j}}{10^{-4}\rm cm^{-3}}\right)
\left(\frac{V}{0.36\,\rm kpc^{3}}\right) 
\left(\frac{\Gamma_{\rm jet} -1}{0.06}\right)\right]^{-\frac{2}{3}}.
\end{aligned}
\end{eqnarray}
In Figure~\ref{B} we plot $B_{\rm min}$ and $B_{\rm min,a}$ (green-dashed line), 
and we  see that $B_{\rm min,a}$ is a very good approximation.
Finally, the hotspot magnetic field required to explain the synchrotron 
emission at 43~GHz is $50 \lesssim B \lesssim 400$~$\mu$G.
We keep  $V$ in Eqs.~(\ref{Beq}) and (\ref{Bmin}) to show that 
$B_{\rm eq}$ and $B_{\rm min,a}$, and
therefore $B_{\rm min}$, increases when $V$ is smaller than $0.36$~kpc$^3$.
This is the case when we take into account that the jet is inclined by an 
angle  $\sim 70^{\circ}$ with the line of sight 
\citep{Boccardi_16, Steenbrugge_08}. In such a case, the real extent of 
the synchrotron emitter is smaller than $l_{\rm obs}$ and therefore 
the emitter volume is smaller than $0.36$~kpc$^3$ \citep{Meisenheimer_89}.

The hotspot magnetic field could also be constrained by modeling the 
(synchrotron self Compton) X-ray emission 
\cite[see e.g.][]{Wright_04,Stawarz_07}. However, we need to know the 
X-ray-emitter volume which is not easy to determine from 
the data in the X-ray domain. 

\section{Cut-off of the synchrotron spectrum}
\label{cutoff}

Diffuse IR (at frequencies $3.798\times10^{13}$ and $6.655\times10^{13}$~Hz) 
and optical ($\nu_{\rm opt} = 5.45\times10^{14}$~Hz) 
emission was detected with the Spitzer and Hubble Space Telescopes, 
respectively \citep{Nilsson_97,Stawarz_07}. 
The very steep IR-to-optical spectral index, $\alpha_{\rm IR-opt} \sim 2.16$, 
indicates that the cut-off of the synchrotron spectrum is  $\nu_{\rm c} < 5\times10^{14}$~Hz. \cite{Stawarz_07} suggested 
that the optical emission  is the low-energy tail of the synchrotron self 
Compton spectrum, 
as in the case of the Cygnus~A northwest secondary hotspot. In such a case,  
$\nu_{\rm c} < 5\times10^{14}$~Hz. The maximum energy of 
non-thermal electrons accelerated at the jet reverse shock is 
$E_{e,\rm max} \sim 0.8$~TeV when $\nu_{\rm c} =  5\times10^{14}$~Hz and
$B = 100$~$\mu$G, as shown in Eq.~(\ref{Ec}).

\subsection{Revising the reigning paradigm}
\label{paradigm}

It is commonly assumed in the literature that $E_{e,\rm max}$ is determined 
by synchrotron losses \cite[e.g.][]{3C445_02Sci}. In such a case, by equating 
the synchrotron cooling time, $t_{\rm synchr} \sim 600/(E_{e,\rm max} B^2)$~s, 
with the acceleration timescale $t_{\rm acc} \sim 20 \mathcal{D}/v_{\rm sh}^2$, 
where the diffusion coefficient is $\mathcal{D}= \lambda c/3$ and
$\lambda$ is the mean-free path, we find that
\begin{equation}
\frac{\mathcal{D}}{\mathcal{D_{\rm Bohm}}} = 
\frac{\lambda}{r_{\rm g}} \sim 2\times10^6
\left(\frac{v_{\rm sh}}{c/3}\right)^2
\left(\frac{\nu_{\rm c}}{5\times 10^{14}\,{\rm Hz}}\right)^{-1}.
\label{DDBohm}
\end{equation}
In Eq.~(\ref{DDBohm})  $\mathcal{D_{\rm Bohm}} = r_{\rm g}c/3$ is the Bohm
diffusion coefficient and $r_{\rm g} = E_{e,\rm max}/(eB)$ is the Larmor radius 
of $E_{e,\rm max}$-electrons (and protons) in a turbulent field $B$. 

\begin{figure}
\includegraphics[angle=0,width=0.49\textwidth]{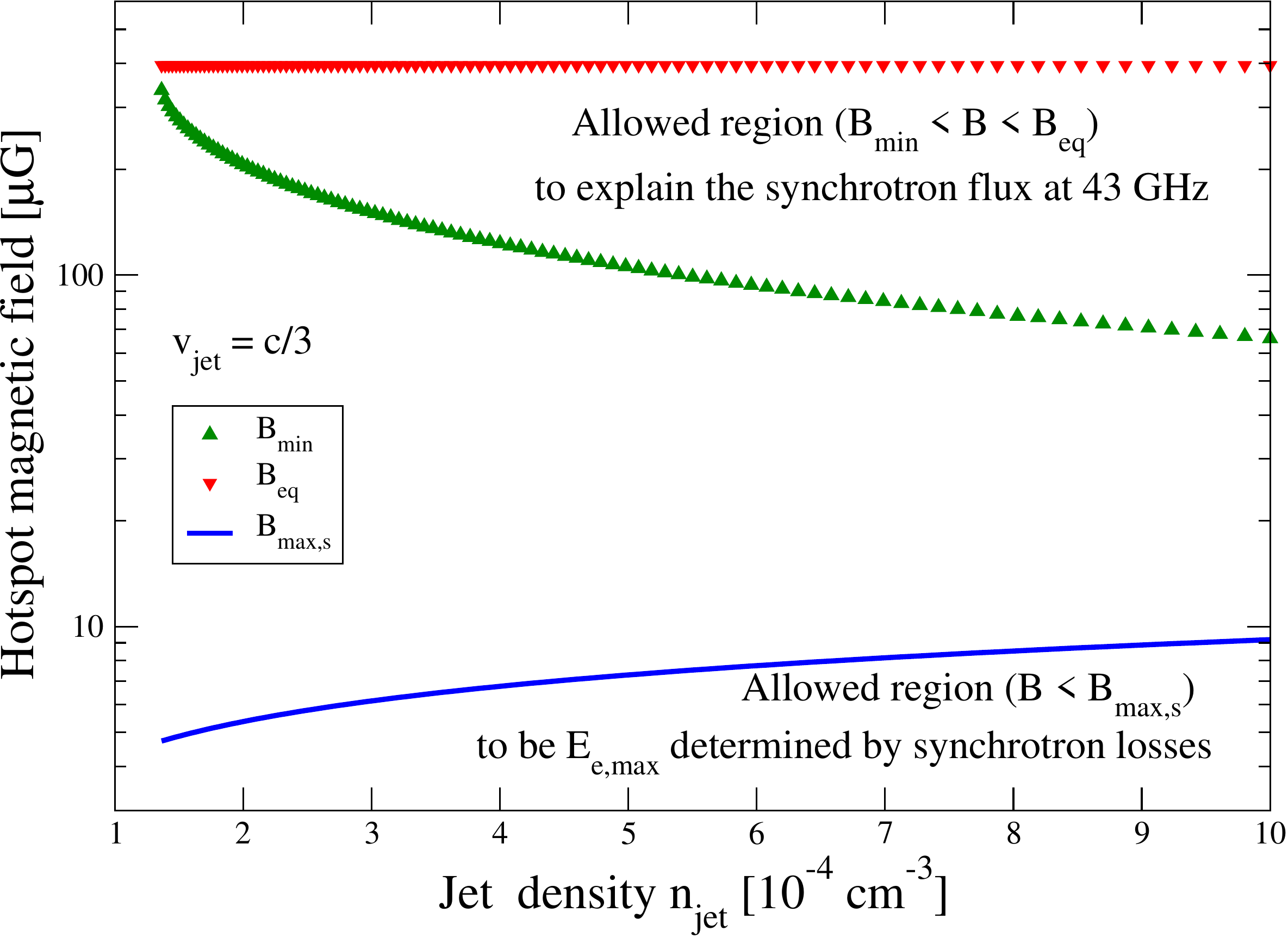}
\caption{Comparison between the magnetic field required to explain the
synchrotron flux at 43~GHz ($B_{\rm min} \le B \le B_{\rm eq}$) and the magnetic
field required to satisfy the condition $s \le c/\omega_{\rm pi}$
($B \le B_{\rm max,s}$).  We can see that $B_{\rm max,s} < B_{\rm min}$ and therefore
the condition $B \le B_{\rm max,s}$ is not satisfied.}
\label{B_const}
\end{figure}

\begin{figure}
\includegraphics[angle=0,width=0.49\textwidth]{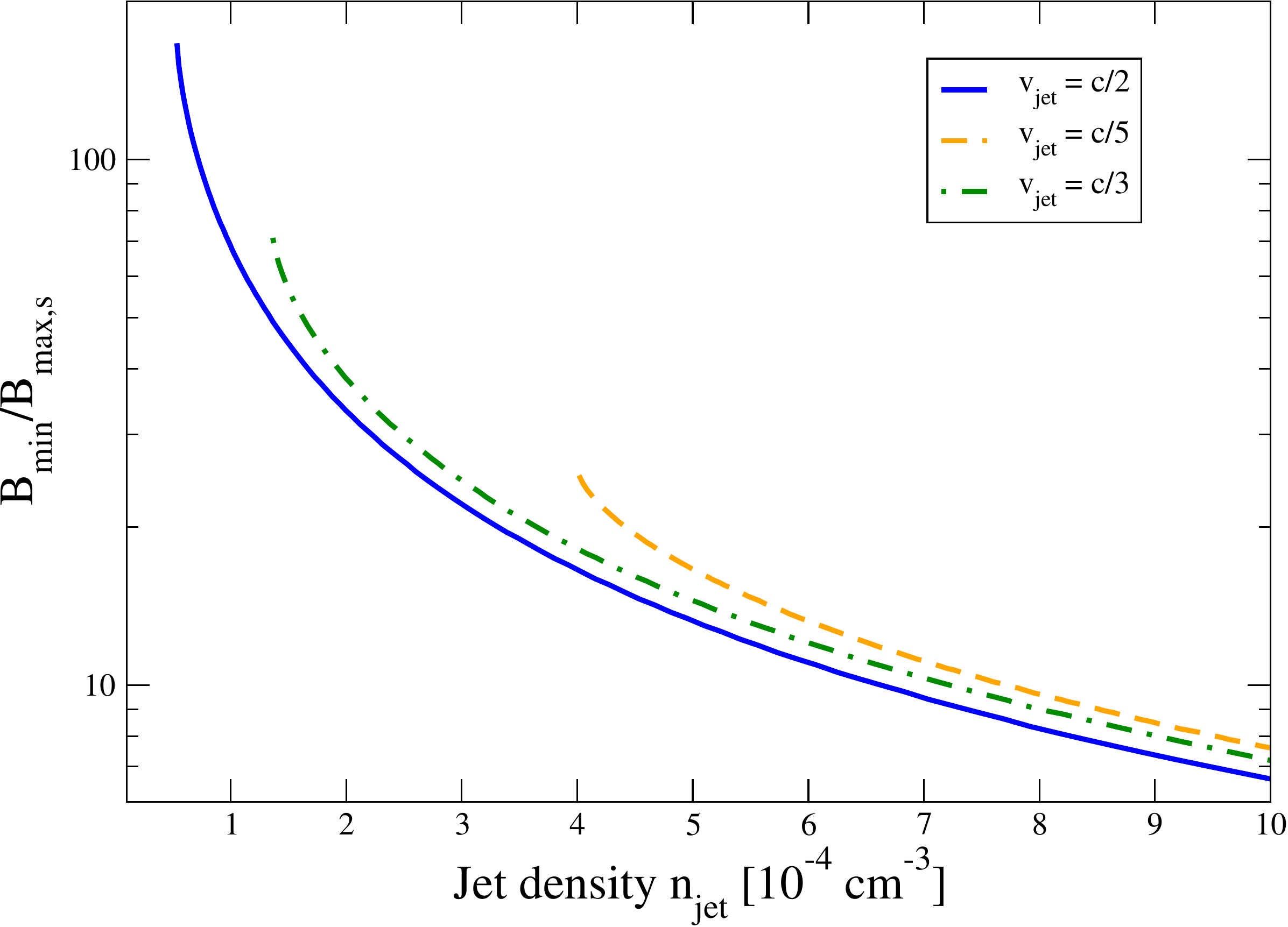}
\caption{Ratio $B_{\rm min}/B_{\rm max,s}$ for $v_{\rm jet} = c/2$ (blue-solid
line), $c/3$ (green-dot-dashed line), and $c/5$ (orange-dashed line).}
\label{Bratio}
\end{figure}

In the small scale turbulence regime $\lambda = r_{\rm g}^2/s$, where $s$ is 
the plasma-turbulence scale-length
\citep[e.g.][]{Ostrowski_02,Kirk_Brian_10,lemoine-pelletier-10,Sironi_13}. 
Therefore, from Eq.~(\ref{DDBohm}), the plasma-turbulence scale-length in the 
"reigning paradigm" is 
\begin{eqnarray}
\begin{aligned}
s &\sim \frac{r_{\rm g}^2}{\lambda} = 
r_{\rm g}\frac{\mathcal{D_{\rm Bohm}}}{\mathcal{D}}\\
&\sim 8.3\times10^6
\left(\frac{\nu_{\rm c}}{5\times 10^{14}\,{\rm Hz}}\right)^{\frac{3}{2}}
\left(\frac{B}{100\,\mu{\rm G}}\right)^{-\frac{3}{2}}
\left(\frac{v_{\rm sh}}{c/3}\right)^{-2}\,{\rm cm}.
\label{s_synchr}
\end{aligned}
\end{eqnarray}
Surprisingly, $s$ is smaller than the ion-skin depth
$c/\omega_{\rm pi}\sim 10^9\,\Gamma_{\rm jet}^{0.5}(n_{\rm jet}/10^{-4}{\rm cm^{-3}})^{-0.5}$~cm
unless $B$ is smaller than 
\begin{equation}\label{Bs}
\frac{B_{\rm max,s}}{\rm \mu G} \sim 2
\left(\frac{\nu_{\rm c}}{5\times 10^{14}\,{\rm Hz}}\right)
\left(\frac{n_{\rm jet}}{10^{-4}\,{\rm cm^{-3}}}\right)^{\frac{1}{3}}
\left(\frac{v_{\rm sh}}{c/3}\right)^{-\frac{4}{3}}.
\end{equation}
(Note that $B_{\rm max,s} \propto \Gamma_{\rm jet}^{-1/3}$, but we
neglect this dependence in Eq.~(\ref{Bs}) given that 
$1.02\le \Gamma_{\rm jet}\le 1.15$ when $c/5 \le v_{\rm jet}\le c/2$.) 
In Figure~\ref{B_const} we plot $B_{\rm max,s}$ (blue-solid line) for the 
case $v_{\rm jet} = c/3$. We see that 
$B_{\rm min}$ is larger than $B_{\rm max,s}$ for all  possible values of 
$n_{\rm jet}$. We mentioned that $5\times 10^{14}$~Hz is the upper-limit 
for the synchrotron spectrum cut-off. In the case that 
$\nu_{\rm c}<5\times 10^{14}$~Hz, $B_{\rm max,s}$ is even smaller than the value 
plotted in Figure~\ref{B_const} whereas $B_{\rm min}$ increases.  Therefore,  
$\nu_{\rm c}<5\times 10^{14}$~Hz enlarges the gap between $B_{\rm min}$  and 
$B_{\rm max,s}$. Note that $B_{\rm min}/B_{\rm max,s}$ also increases when we 
consider an emission volume (at 43~GHz) smaller than $0.36$~kpc$^3$ as a 
consequence of the jet inclination angle \cite[e.g.][]{Meisenheimer_89}.
In Figure~\ref{Bratio} we plot $B_{\rm min}/B_{\rm max,s}$ for the
cases $v_{\rm jet} = c/2$ (blue-solid line),  $v_{\rm jet} = c/3$ 
(green-dot-dashed line), and $c/5$ (orange-dashed line). We can see that
$B_{\rm min} > B_{\rm max,s}$ for all possible values of $v_{\rm jet}$ and 
$n_{\rm jet}$.

Hence we  show that $B$ is larger than $B_{\rm max,s}$ for a large range
of parameters  and therefore $E_{e,\rm max}$ 
cannot be determined by synchrotron cooling in the primary hotspot of
Cygnus A, in disagreement with the standard assumption as was pointed out
by \cite{cutoff_nrh}.  Note that to reach this conclusion we have  only used
well resolved radio emission at 43~GHz and the requirement  
$s > c/\omega_{\rm pi}$. In the next section we explore a more fundamental
limit to constrain $E_{e,\rm max}$.

\section{The case of perpendicular shocks}

The maximum energy is ultimately constrained by the ability to scatter 
particles back and forth across the shock, and this depends on the geometry 
of the magnetic field (i.e. the angle between the field vector and the shock 
normal). In this section we consider the case of perpendicular shocks, given 
that relativistic shocks are characteristically quasi-perpendicular. 
Note however that shocks moving at $v_{\rm sh}\sim c/3$ are mildly 
relativistic and therefore they may not be strictly perpendicular.
Unfortunately, it is not possible to determine the geometry of the magnetic 
field in the reverse shock downstream region using the  polarization data 
available in the literature. 

\subsection{Electrons' maximum energy determined by the diffusion condition}
\label{s_optical}

\begin{figure}
\includegraphics[width=0.49\textwidth]{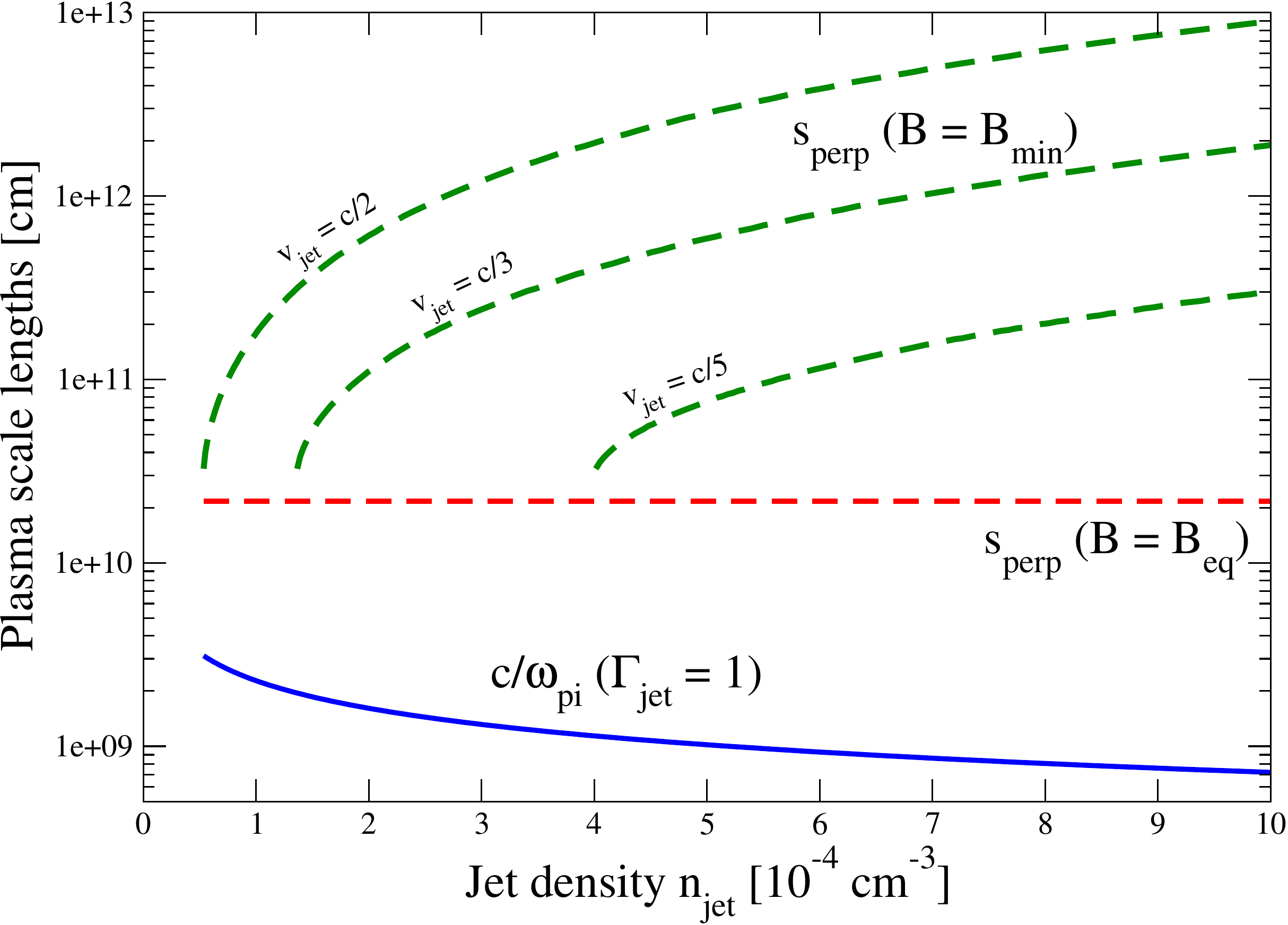}
\caption{Magnetic turbulence scale-length 
for the case of perpendicular shocks ($s_{\perp}$ -  dashed lines)
and ion-skin-depth $c/\omega_{\rm pi}$ (blue-solid line). 
\label{scales}}
\end{figure}

To accelerate particles  up to an energy $E_{e,\rm max}$ in perpendicular shocks, 
the mean-free path in turbulent magnetic field in the shock downstream region, 
$\lambda_{\rm d} \sim (E_{e,\rm max}/eB)^2/s$
has to be smaller than Larmor radius in $B_{\rm jd}$ in order to avoid the
particles following the $B_{\rm jd}$-helical orbits and cross-field diffusion 
ceasing \citep{Kirk_Brian_10,lemoine-pelletier-10,Sironi_13,Brian_14}. 
The condition $\lambda_{\rm d} \lesssim r_{\rm g0}$, where 
$r_{\rm g0} = E_{e,\rm max}/(eB_{\rm jd})$ is the Larmor radius in the ordered 
(and compressed) 
field $B_{\rm jd} \sim 4 B_{\rm j}$, where $B_{\rm j}$ is the jet magnetic field,  is marginally satisfied when the magnetic-turbulence 
scale-length is $s = s_{\perp}$, where
\begin{eqnarray}\label{s_nrh}
\begin{aligned}
s_{\perp} &=\frac{E_{e,\rm max}}{eB}\left(\frac{4B_{\rm j}}{B}\right) \\
&\sim 6.7\times10^{11}
\left(\frac{\nu_{\rm c}}{5\times 10^{14}\,{\rm Hz}}\right)^{\frac{1}{2}}
\left(\frac{B_{\rm j}}{\rm \mu G}\right)
\left(\frac{B}{\rm 100\mu G}\right)^{-\frac{5}{2}}\,{\rm cm}.    
\end{aligned}
\end{eqnarray}
In Figure~\ref{scales} we plot $s_{\perp}$ for the cases of $B = B_{\rm eq}$ 
(red-dotted line) and $B = B_{\rm min}$ (green-dotted lines) and fixing 
$B_{\rm j}=1\,\mu$G. We plot also  $c/\omega_{\rm pi} (\propto n_{\rm jet}^{-0.5})$.
Note that  $s_{\perp} > c/\omega_{\rm pi}$ which indicates that the magnetic 
field is probably not generated by the Weibel instability 
(that has a characteristic scale length of  $c/\omega_{\rm pi}$).

\subsection{NRH instabilities in perpendicular shocks}
\label{nrh}

Turbulence on a scale greater than $c/\omega_{\rm pi}$ may be excited through
the NRH instability, which can grow until $s$ reaches the Larmor radius of the
highest energy CR driving the instability \citep{Bell_04,Tony_05}. 
Since the scattering rate is proportional to $E^{-2}$ in given small scale 
turbulence, the distance over which CR currents are anisotropised downstream
of the shock is proportional to $E^2$. Hence the higher energy CR have more 
time to drive the NRH instability, and CR with energy $E_{\rm e,max}$
are predominantly responsible for generating the turbulence unless the CR 
spectrum is unusually steep ($p>3$). As explained above, the maximum CR energy 
$E_{\rm e,max}$ is that of CR whose anistropy decays over a distance equal to 
their Larmor radius  in the ordered component of the downstream magnetic 
field \citep{Tony_17}

We now discuss whether $E_{\rm e,max}$-CRs have sufficient energy density to 
amplify the magnetic field. To amplify the magnetic field via the NRH 
instability in a perpendicular shock, the turbulent field has to grow through 
around 10 e-foldings at the maximum growth rate $\Gamma_{\rm max}$ 
\citep{Tony_escape_13}.  The time available for the instability to grow is 
$t_{\perp} = r_{\rm g0}/v_{\rm d}$ during which the plasma  flows through a 
distance $r_{\rm g0}$ in the downstream region
at velocity $v_{\rm d}\sim v_{\rm sh}/4$.  Therefore, the condition for 
magnetic field amplification by the NRH instability in  perpendicular shocks is
$\Gamma_{\rm max}t_{\perp} > 10$. In perpendicular shocks where both the CR 
current ${\bf j_{\rm CR}}$ and 
${\bf B_{\rm jd}}$ are in the plane of the shock and
orthogonal to each other, $\Gamma_{\rm max}$ is similar to the 
linear growth rate in parallel shocks, as shown by \cite{Riquelme_10} and
\cite{James_17},  and in
agreement with the dispersion relation derived by \cite{Tony_05}.   
Therefore,  in perpendicular geometry, 
$\Gamma_{\rm max}\sim (j_{\rm CR}/c) \sqrt{\pi/\rho_{\rm jet}}$, 
where $\rho_{\rm jet} = m_p n_{\rm jet}$.  

The current density carried by $E_{\rm e,max}$-CRs is 
$j_{\rm e,max} = \eta_{\rm e,max} U_{\rm kin} c\,e/E_{\rm e,max}$,
where $ \eta_{\rm e,max} =1$ notionally represents the condition in which the 
CR electron number density at energy $E _{\rm e,max}$
is equal to $U_{\rm kin}/E_{\rm e,max} $ and the CR drift along 
the shock surface at velocity $c$.
Allowing for compression of the mass density and magnetic field by a factor 
of four at the shock, the condition $\Gamma_{\max} t_\perp >10$ leads to a 
lower limit on $\eta _{\rm e,max}$:
$\eta _{e,\rm max} > \eta_{\rm min} = 80/M_{\rm A}$
where $M_{\rm A}=v_{\rm sh}/v_{\rm A}$ is the Alfven Mach number of the jet at the 
termination shock and  $v_{\rm A}=B_{\rm j}/\sqrt{4 \pi \rho _{\rm jet}}$, giving
\begin{equation}
M_{\rm A}= 1400 \left(\frac{v_{\rm sh}}{c/3}\right) 
\left(\frac{B_{\rm j}}{\rm \mu G}\right)^{-1}
\left(\frac{n_{\rm jet}}{10^{-4}\,{\rm cm^{-3}}}\right)^{\frac{1}{2}}
\end{equation}
and therefore 
\begin{equation}\label{eta_min}
\eta_{\rm min} = 0.057  \left(\frac{v_{\rm sh}}{c/3}\right)^{-1} 
\left(\frac{B_{\rm j}}{\rm \mu G}\right)
\left(\frac{n_{\rm jet}}{10^{-4}\,{\rm cm^{-3}}}\right)^{-\frac{1}{2}}.
\end{equation}
The ordered magnetic field $B_{\rm j}$  in the termination region
of AGN jets is unknown, but
values lower than $1 \mu {\rm G}$ are reasonable 
considering the lateral expansion of the jet during propagation from its 
origin in the active galactic nucleus.
It appears that the CR current is sufficient to  drive the NRH instability, 
but the margins are tight, CR acceleration to energy $E _{\rm e,max}$ must be 
efficient, and the jet magnetic field must be small.

In order to check whether these conditions are satisfied in the primary hotspot
of Cygnus~A, we consider that non-thermal protons are accelerated in the
jet reverse shock following a power-law energy distribution with the same 
index as non-thermal electrons ($p=2.44$). In such a case, the energy density 
in  $E_{e,\rm max}$-protons is $U_{e,\rm max} = K_p E_{e,\rm max}^{2-p}$, where $K_p$ 
is the normalization constant of the energy distribution. Considering that 
$U_e = U_p$ (see Section~\ref{B_43})  we find 
$K_p = U_e (p-2)/E_{p,\rm min}^{2-p}$ where $E_{p,\rm min}$ 
is the minimum energy of non-thermal protons. By setting  $E_{p,\rm min}=1$~GeV 
we find that the acceleration efficiency of $E_{e,\rm max}$-protons is
\begin{eqnarray}\label{eta_nrh}
\begin{aligned}
\eta_{e,\rm max} &\equiv \frac{U_{e,\rm max}}{U_{\rm kin}}\sim 0.44\left(\frac{U_e}{U_{\rm kin}}\right)
\left(\frac{E_{e,\rm max}}{\rm GeV}\right)^{-0.44}\\
&\sim 0.07\left[\left(\frac{\nu_{\rm c}}{5\times10^{14}\,{\rm Hz}}\right)
\left(\frac{\nu_{\rm min}}{0.1\,{\rm GHz}}\right)\right]^{-0.22}
\left(\frac{B}{\rm 100\,\mu G}\right)^{-1.28}\\
&\times\left[\left(\frac{n_{\rm jet}}{10^{-4}\,{\rm cm^{-3}}}\right)
\left(\frac{\Gamma_{\rm jet}-1}{0.06}\right)
\left(\frac{V}{0.36\,\rm kpc^3}\right)\right]^{-1}
\end{aligned}
\end{eqnarray}
Therefore, to satisfy the condition $\eta_{e,\rm max}>\eta_{\rm min}$ 
(Eq.~\ref{eta_min}) for efficient magnetic field amplification by the 
NRH-instability in a perpendicular shock, the jet (unperturbed) magnetic 
field has to be 
\begin{eqnarray}\label{B_nrh}
\begin{aligned}
\left(\frac{B_{\rm j}}{\rm \mu G}\right) & < 1.2
\left(\frac{n_{\rm jet}}{10^{-4}\,{\rm cm^{-3}}}\right)^{-\frac{1}{2}}
\left(\frac{B}{\rm 100\,\mu G}\right)^{-1.28}
\left(\frac{\Gamma_{\rm jet}-1}{0.06}\right)^{-1},
\end{aligned}
\end{eqnarray}
when $V$, $\nu_{\rm c}$, and $\nu_{\rm min}$
take the values in Table~\ref{tab}, and $v_{\rm sh} \sim c/3$.
In such a case, and assuming that the shock is quasi-perpendicular,
$E_{e,\rm max}$-CRs have sufficient energy density to 
generate NRH-turbulence on scale $s_{\perp}$ and
amplify the magnetic field by a factor $B/B_{\rm j} \sim 100$ in the 
primary hotspot of Cygnus~A.

\section{Conclusions}
\label{conclusion}

We study diffusive shock acceleration and magnetic field amplification in 
the northwest primary hostspot in   
Cygnus~A. We focus on the well resolved region downstream
of the jet reverse shock where most of the synchrotron radiation is emitted. 
By considering  the synchrotron  flux at 
43~GHz we determine that the jet density has to be larger  than 
$\sim 5\times10^{-5}$~cm$^{-3}$ and the hotspot magnetic field is 
$50\lesssim B \lesssim 400$~$\mu$G (when the energy density in non-thermal 
protons is the same as in non-thermal electrons, i.e. $a=1$, and 
$c/5 < v_{\rm jet} < c/2$).
The cut-off of the synchrotron spectrum is at 
$\nu_{\rm c}\lesssim5\times10^{14}$~Hz, implying that the maximum energy of 
electrons accelerated in the hotspots is $E_{e,\rm max}< 1$~TeV. By setting 
the magnetic-turbulence scale-length $s$ larger than the ion-skin depth 
$c/\omega_{\rm pi}$ (in the small-scale turbulence regime) we find that the 
magnetic field required to be $E_{e,\rm max}$ determined by synchrotron cooling 
is smaller than the field required to explain the synchrotron emission at 
43~GHz. Therefore, we conclude that $E_{e,\rm max}$ is not constrained by 
synchrotron cooling, as traditionally assumed. 

The maximum energy $E_{e,\rm max}$ is ultimately determined by the scattering 
process.  By assuming  that the shock is quasi-perpendicular,
particles cannot diffuse further than a distance $r_{\rm g0}$ 
downstream of the shock, i.e. $\lambda_{\rm d}< r_{\rm g0}$. To satisfy this 
condition, the magnetic turbulence scale-length has to be larger than
$\sim 2\times10^{10}$~cm, that is $\sim 10\, c/\omega_{\rm pi}$ 
(see Fig.~\ref{scales}), and therefore $B$ is probably not amplified 
by the Weibel  turbulence. 

On the other hand, the NRH instability amplify the magnetic field on scales 
larger than $c/\omega_{\rm pi}$ and we show that  NRH-modes
generated by CRs with energies $E_{e,\rm max}$ can grow fast enough 
to amplify the jet magnetic field 
from $\sim$1 to 100~$\mu$G and accelerate particles up to energies
$E_{e,\rm max}\sim 0.8$~TeV observed in the primary hotspot of Cygnus~A 
radiogalaxy.
The advantage of magnetic turbulence being generated by CRs current is that
$B$ persists over long distances downstream of the 
shock, and therefore particles accelerated very near the shock can emit
synchrotron radiation far downstream. 

Finally, if $E_{e,\rm max}$ is determined by the diffusion condition in a 
perpendicular shock, the same limit applies to protons and therefore the 
maximum energy of ions is also $\sim 0.8$~TeV. As a consequence, relativistic 
shocks in the termination region of FR~II jets are poor cosmic ray accelerators.

\section*{Acknowledgements}

The authors thank the anonymous referee for a constructive report.  
The authors thank  S.~Pyrzas for providing Figure~3 in \cite{Pyrzas_15}, 
and Alexandre Marcowith and Robert Laing for useful comments. 
The research leading to this article has received funding
from the European Research Council under the European
Community's Seventh Framework Programme (FP7/2007-2013)/ERC grant agreement 
no. 247039. We acknowledge support from the UK
Science and Technology Facilities Council under grants 
ST/K00106X and ST/N000919/1.

\bibliographystyle{mnras}
\bibliography{biblio_CRs} 

\begin{thebibliography}{}
\makeatletter
\relax
\def\mn@urlcharsother{\let\do\@makeother \do\$\do\&\do\#\do\^\do\_\do\%\do\~}
\def\mn@doi{\begingroup\mn@urlcharsother \@ifnextchar [ {\mn@doi@}
  {\mn@doi@[]}}
\def\mn@doi@[#1]#2{\def\@tempa{#1}\ifx\@tempa\@empty \href
  {http://dx.doi.org/#2} {doi:#2}\else \href {http://dx.doi.org/#2} {#1}\fi
  \endgroup}
\def\mn@eprint#1#2{\mn@eprint@#1:#2::\@nil}
\def\mn@eprint@arXiv#1{\href {http://arxiv.org/abs/#1} {{\tt arXiv:#1}}}
\def\mn@eprint@dblp#1{\href {http://dblp.uni-trier.de/rec/bibtex/#1.xml}
  {dblp:#1}}
\def\mn@eprint@#1:#2:#3:#4\@nil{\def\@tempa {#1}\def\@tempb {#2}\def\@tempc
  {#3}\ifx \@tempc \@empty \let \@tempc \@tempb \let \@tempb \@tempa \fi \ifx
  \@tempb \@empty \def\@tempb {arXiv}\fi \@ifundefined
  {mn@eprint@\@tempb}{\@tempb:\@tempc}{\expandafter \expandafter \csname
  mn@eprint@\@tempb\endcsname \expandafter{\@tempc}}}

\bibitem[\protect\citeauthoryear{{Araudo}, {Bell}, {Crilly}  \&
  {Blundell}}{{Araudo} et~al.}{2016}]{cutoff_nrh}
{Araudo} A.~T.,  {Bell} A.~R.,  {Crilly} A.,   {Blundell} K.~M.,  2016, \mn@doi
  [\mnras] {10.1093/mnras/stw1204}, \href
  {http://adsabs.harvard.edu/abs/2016MNRAS.tmp..889A} {}

\bibitem[\protect\citeauthoryear{{Bell}}{{Bell}}{2004}]{Bell_04}
{Bell} A.~R.,  2004, \mn@doi [\mnras] {10.1111/j.1365-2966.2004.08097.x}, \href
  {http://adsabs.harvard.edu/abs/2004MNRAS.353..550B} {353, 550}

\bibitem[\protect\citeauthoryear{{Bell}}{{Bell}}{2005}]{Tony_05}
{Bell} A.~R.,  2005, \mn@doi [\mnras] {10.1111/j.1365-2966.2005.08774.x}, \href
  {http://adsabs.harvard.edu/abs/2005MNRAS.358..181B} {358, 181}

\bibitem[\protect\citeauthoryear{{Bell}, {Schure}, {Reville}  \&
  {Giacinti}}{{Bell} et~al.}{2013}]{Tony_escape_13}
{Bell} A.~R.,  {Schure} K.~M.,  {Reville} B.,   {Giacinti} G.,  2013, \mn@doi
  [\mnras] {10.1093/mnras/stt179}, \href
  {http://adsabs.harvard.edu/abs/2013MNRAS.431..415B} {431, 415}

\bibitem[\protect\citeauthoryear{{Bell}, {Araudo}, {Matthews}  \&
  {Blundell}}{{Bell} et~al.}{2017}]{Tony_17}
{Bell} A.,  {Araudo} A.,  {Matthews} J.,   {Blundell} K.,  2017, preprint,
  \href {http://adsabs.harvard.edu/abs/2017arXiv170907793B} {} (\mn@eprint
  {arXiv} {1709.07793})

\bibitem[\protect\citeauthoryear{{Boccardi}, {Krichbaum}, {Bach}, {Mertens},
  {Ros}, {Alef}  \& {Zensus}}{{Boccardi} et~al.}{2016}]{Boccardi_16}
{Boccardi} B.,  {Krichbaum} T.~P.,  {Bach} U.,  {Mertens} F.,  {Ros} E.,
  {Alef} W.,   {Zensus} J.~A.,  2016, \mn@doi [\aap]
  {10.1051/0004-6361/201526985}, \href
  {http://adsabs.harvard.edu/abs/2016A%26A...585A..33B} {585, A33}

\bibitem[\protect\citeauthoryear{{Brunetti}, {Mack}, {Prieto}  \&
  {Varano}}{{Brunetti} et~al.}{2003}]{Brunetti_03}
{Brunetti} G.,  {Mack} K.-H.,  {Prieto} M.~A.,   {Varano} S.,  2003, \mn@doi
  [\mnras] {10.1046/j.1365-8711.2003.07185.x}, \href
  {http://adsabs.harvard.edu/abs/2003MNRAS.345L..40B} {345, L40}

\bibitem[\protect\citeauthoryear{{Carilli}, {Perley}, {Dreher}  \&
  {Leahy}}{{Carilli} et~al.}{1991}]{Carilli_91}
{Carilli} C.~L.,  {Perley} R.~A.,  {Dreher} J.~W.,   {Leahy} J.~P.,  1991,
  \mn@doi [\apj] {10.1086/170813}, \href
  {http://adsabs.harvard.edu/abs/1991ApJ...383..554C} {383, 554}

\bibitem[\protect\citeauthoryear{{Dreher}, {Carilli}  \& {Perley}}{{Dreher}
  et~al.}{1987}]{polarization}
{Dreher} J.~W.,  {Carilli} C.~L.,   {Perley} R.~A.,  1987, \mn@doi [\apj]
  {10.1086/165229}, \href {http://adsabs.harvard.edu/abs/1987ApJ...316..611D}
  {316, 611}

\bibitem[\protect\citeauthoryear{{Hillas}}{{Hillas}}{1984}]{Hillas}
{Hillas} A.~M.,  1984, \mn@doi [\araa] {10.1146/annurev.aa.22.090184.002233},
  \href {http://adsabs.harvard.edu/abs/1984ARA%26A..22..425H} {22, 425}

\bibitem[\protect\citeauthoryear{{Kirk} \& {Reville}}{{Kirk} \&
  {Reville}}{2010}]{Kirk_Brian_10}
{Kirk} J.~G.,  {Reville} B.,  2010, \mn@doi [\apjl]
  {10.1088/2041-8205/710/1/L16}, \href
  {http://adsabs.harvard.edu/abs/2010ApJ...710L..16K} {710, L16}

\bibitem[\protect\citeauthoryear{{Lagage} \& {Cesarsky}}{{Lagage} \&
  {Cesarsky}}{1983}]{Lagage-Cesarsky}
{Lagage} P.~O.,  {Cesarsky} C.~J.,  1983, \aap, \href
  {http://adsabs.harvard.edu/abs/1983A%26A...125..249L} {125, 249}

\bibitem[\protect\citeauthoryear{{Leahy}, {Muxlow}  \& {Stephens}}{{Leahy}
  et~al.}{1989}]{Leahy_89}
{Leahy} J.~P.,  {Muxlow} T.~W.~B.,   {Stephens} P.~W.,  1989, \mn@doi [\mnras]
  {10.1093/mnras/239.2.401}, \href
  {http://adsabs.harvard.edu/abs/1989MNRAS.239..401L} {239, 401}

\bibitem[\protect\citeauthoryear{{Lemoine} \& {Pelletier}}{{Lemoine} \&
  {Pelletier}}{2010}]{lemoine-pelletier-10}
{Lemoine} M.,  {Pelletier} G.,  2010, \mn@doi [\mnras]
  {10.1111/j.1365-2966.2009.15869.x}, \href
  {http://adsabs.harvard.edu/abs/2010MNRAS.402..321L} {402, 321}

\bibitem[\protect\citeauthoryear{{Mack}, {Prieto}, {Brunetti}  \&
  {Orienti}}{{Mack} et~al.}{2009}]{Mack_09}
{Mack} K.-H.,  {Prieto} M.~A.,  {Brunetti} G.,   {Orienti} M.,  2009, \mn@doi
  [\mnras] {10.1111/j.1365-2966.2008.14081.x}, \href
  {http://adsabs.harvard.edu/abs/2009MNRAS.392..705M} {392, 705}

\bibitem[\protect\citeauthoryear{{Matthews}, {Bell}, {Blundell}  \&
  {Araudo}}{{Matthews} et~al.}{2017}]{James_17}
{Matthews} J.~H.,  {Bell} A.~R.,  {Blundell} K.~M.,   {Araudo} A.~T.,  2017,
  \mn@doi [\mnras] {10.1093/mnras/stx905}, \href
  {http://adsabs.harvard.edu/abs/2017MNRAS.469.1849M} {469, 1849}

\bibitem[\protect\citeauthoryear{{McKean} et~al.,}{{McKean}
  et~al.}{2016}]{CygA_gammamin}
{McKean} J.~P.,  et~al., 2016, \mn@doi [\mnras] {10.1093/mnras/stw2105}, \href
  {http://adsabs.harvard.edu/abs/2016MNRAS.463.3143M} {463, 3143}

\bibitem[\protect\citeauthoryear{{Meisenheimer} \& {Heavens}}{{Meisenheimer} \&
  {Heavens}}{1986}]{3c273-Natur}
{Meisenheimer} K.,  {Heavens} A.~F.,  1986, \mn@doi [\nat] {10.1038/323419a0},
  \href {http://adsabs.harvard.edu/abs/1986Natur.323..419M} {323, 419}

\bibitem[\protect\citeauthoryear{{Meisenheimer}, {Roser}, {Hiltner}, {Yates},
  {Longair}, {Chini}  \& {Perley}}{{Meisenheimer}
  et~al.}{1989}]{Meisenheimer_89}
{Meisenheimer} K.,  {Roser} H.-J.,  {Hiltner} P.~R.,  {Yates} M.~G.,  {Longair}
  M.~S.,  {Chini} R.,   {Perley} R.~A.,  1989, \aap, \href
  {http://adsabs.harvard.edu/abs/1989A%26A...219...63M} {219, 63}

\bibitem[\protect\citeauthoryear{{Nilsson}, {Valtonen}, {Jones}, {Saslaw}  \&
  {Lehto}}{{Nilsson} et~al.}{1997}]{Nilsson_97}
{Nilsson} K.,  {Valtonen} M.~J.,  {Jones} L.~R.,  {Saslaw} W.~C.,   {Lehto}
  H.~J.,  1997, \aap, \href
  {http://adsabs.harvard.edu/abs/1997A%26A...324..888N} {324, 888}

\bibitem[\protect\citeauthoryear{{Norman}, {Melrose}  \& {Achterberg}}{{Norman}
  et~al.}{1995}]{Norman_95}
{Norman} C.~A.,  {Melrose} D.~B.,   {Achterberg} A.,  1995, \mn@doi [\apj]
  {10.1086/176465}, \href {http://adsabs.harvard.edu/abs/1995ApJ...454...60N}
  {454, 60}

\bibitem[\protect\citeauthoryear{{Orienti}, {Brunetti}, {Nagai}, {Paladino},
  {Mack}  \& {Prieto}}{{Orienti} et~al.}{2017}]{Orienti_17}
{Orienti} M.,  {Brunetti} G.,  {Nagai} H.,  {Paladino} R.,  {Mack} K.-H.,
  {Prieto} M.~A.,  2017, preprint, \href
  {http://adsabs.harvard.edu/abs/2017arXiv170506465O} {} (\mn@eprint {arXiv}
  {1705.06465})

\bibitem[\protect\citeauthoryear{{Ostrowski} \& {Bednarz}}{{Ostrowski} \&
  {Bednarz}}{2002}]{Ostrowski_02}
{Ostrowski} M.,  {Bednarz} J.,  2002, \mn@doi [\aap]
  {10.1051/0004-6361:20021173}, \href
  {http://adsabs.harvard.edu/abs/2002A%26A...394.1141O} {394, 1141}

\bibitem[\protect\citeauthoryear{{Owen}, {Ledlow}, {Morrison}  \&
  {Hill}}{{Owen} et~al.}{1997}]{Owen_97}
{Owen} F.~N.,  {Ledlow} M.~J.,  {Morrison} G.~E.,   {Hill} J.~M.,  1997,
  \mn@doi [\apjl] {10.1086/310908}, \href
  {http://adsabs.harvard.edu/abs/1997ApJ...488L..15O} {488, L15}

\bibitem[\protect\citeauthoryear{{Perlman}, {Georganopoulos}, {May}  \&
  {Kazanas}}{{Perlman} et~al.}{2010}]{Perlman_10}
{Perlman} E.~S.,  {Georganopoulos} M.,  {May} E.~M.,   {Kazanas} D.,  2010,
  \mn@doi [\apj] {10.1088/0004-637X/708/1/1}, \href
  {http://adsabs.harvard.edu/abs/2010ApJ...708....1P} {708, 1}

\bibitem[\protect\citeauthoryear{{Prieto}, {Brunetti}  \& {Mack}}{{Prieto}
  et~al.}{2002}]{3C445_02Sci}
{Prieto} M.~A.,  {Brunetti} G.,   {Mack} K.-H.,  2002, \mn@doi [Science]
  {10.1126/science.1075990}, \href
  {http://adsabs.harvard.edu/abs/2002Sci...298..193P} {298, 193}

\bibitem[\protect\citeauthoryear{{Pyrzas}, {Steenbrugge}  \&
  {Blundell}}{{Pyrzas} et~al.}{2015}]{Pyrzas_15}
{Pyrzas} S.,  {Steenbrugge} K.~C.,   {Blundell} K.~M.,  2015, \mn@doi [\aap]
  {10.1051/0004-6361/201425061}, \href
  {http://adsabs.harvard.edu/abs/2015A%26A...574A..30P} {574, A30}

\bibitem[\protect\citeauthoryear{{Rachen} \& {Biermann}}{{Rachen} \&
  {Biermann}}{1993}]{Rachen_93}
{Rachen} J.~P.,  {Biermann} P.~L.,  1993, \aap, \href
  {http://adsabs.harvard.edu/abs/1993A%26A...272..161R} {272, 161}

\bibitem[\protect\citeauthoryear{{Reville} \& {Bell}}{{Reville} \&
  {Bell}}{2014}]{Brian_14}
{Reville} B.,  {Bell} A.~R.,  2014, \mn@doi [\mnras] {10.1093/mnras/stu088},
  \href {http://adsabs.harvard.edu/abs/2014MNRAS.439.2050R} {439, 2050}

\bibitem[\protect\citeauthoryear{{Riquelme} \& {Spitkovsky}}{{Riquelme} \&
  {Spitkovsky}}{2010}]{Riquelme_10}
{Riquelme} M.~A.,  {Spitkovsky} A.,  2010, \mn@doi [\apj]
  {10.1088/0004-637X/717/2/1054}, \href
  {http://adsabs.harvard.edu/abs/2010ApJ...717.1054R} {717, 1054}

\bibitem[\protect\citeauthoryear{{Sironi}, {Spitkovsky}  \& {Arons}}{{Sironi}
  et~al.}{2013}]{Sironi_13}
{Sironi} L.,  {Spitkovsky} A.,   {Arons} J.,  2013, \mn@doi [\apj]
  {10.1088/0004-637X/771/1/54}, \href
  {http://adsabs.harvard.edu/abs/2013ApJ...771...54S} {771, 54}

\bibitem[\protect\citeauthoryear{{Stawarz}, {Cheung}, {Harris}  \&
  {Ostrowski}}{{Stawarz} et~al.}{2007}]{Stawarz_07}
{Stawarz} {\L}.,  {Cheung} C.~C.,  {Harris} D.~E.,   {Ostrowski} M.,  2007,
  \mn@doi [\apj] {10.1086/517966}, \href
  {http://adsabs.harvard.edu/abs/2007ApJ...662..213S} {662, 213}

\bibitem[\protect\citeauthoryear{{Steenbrugge} \& {Blundell}}{{Steenbrugge} \&
  {Blundell}}{2008}]{Steenbrugge_08}
{Steenbrugge} K.~C.,  {Blundell} K.~M.,  2008, \mn@doi [\mnras]
  {10.1111/j.1365-2966.2007.12665.x}, \href
  {http://adsabs.harvard.edu/abs/2008MNRAS.388.1457S} {388, 1457}

\bibitem[\protect\citeauthoryear{{Tingay}, {Lenc}, {Brunetti}  \&
  {Bondi}}{{Tingay} et~al.}{2008}]{Tingay_08}
{Tingay} S.~J.,  {Lenc} E.,  {Brunetti} G.,   {Bondi} M.,  2008, \mn@doi [\aj]
  {10.1088/0004-6256/136/6/2473}, \href
  {http://adsabs.harvard.edu/abs/2008AJ....136.2473T} {136, 2473}

\bibitem[\protect\citeauthoryear{{Wilson}, {Young}  \& {Shopbell}}{{Wilson}
  et~al.}{2000}]{Wilson_00}
{Wilson} A.~S.,  {Young} A.~J.,   {Shopbell} P.~L.,  2000, \mn@doi [\apjl]
  {10.1086/317293}, \href {http://adsabs.harvard.edu/abs/2000ApJ...544L..27W}
  {544, L27}

\bibitem[\protect\citeauthoryear{{Wilson}, {Smith}  \& {Young}}{{Wilson}
  et~al.}{2006}]{Wilson_06}
{Wilson} A.~S.,  {Smith} D.~A.,   {Young} A.~J.,  2006, \mn@doi [\apjl]
  {10.1086/504108}, \href {http://adsabs.harvard.edu/abs/2006ApJ...644L...9W}
  {644, L9}

\bibitem[\protect\citeauthoryear{{Wright} \& {Birkinshaw}}{{Wright} \&
  {Birkinshaw}}{2004}]{Wright_04}
{Wright} M.~C.~H.,  {Birkinshaw} M.,  2004, \mn@doi [\apj] {10.1086/423483},
  \href {http://adsabs.harvard.edu/abs/2004ApJ...614..115W} {614, 115}

\makeatother
\end{thebibliography}


\bsp	
\label{lastpage}
\end{document}